\documentclass{nsr}

\usepackage{amsmath,graphicx,array}
\usepackage{dcolumn,soul}%

\usepackage{amsthm}
\usepackage[figuresright]{rotating}%
\usepackage{algorithm, algorithmicx, algpseudocode}
\usepackage{listings}%
\usepackage{hyperref}
\usepackage{xurl}
\usepackage{comment}
\usepackage{pdfpages}

\makeatletter
\def\uns{\ifmmode\,\else$\,$\fi}%

\makeatother
 
\jvol{XX}
\jnum{X}
\jyear{Year}
\doi{10.1093/nsr/XXXX}
\received{XX XX Year}
\revised{XX XX Year}
\accepted{XX XX Year}

\begin{document}

\dhead{RESEARCH ARTICLE}

\subhead{PHYSICS}

\title{Cygnus X-3: A variable petaelectronvolt gamma-ray source}
\author{The LHAASO Collaboration$^{\dagger}$,*}

\authornote{\textbf{Corresponding authors.} Emails: caozh@ihep.ac.cn; licong@ihep.ac.cn; jiesh.wang@gmail.com; zjn@shao.ac.cn;
Felix.Aharonian@mpi-hd.mpg.de}

\authornote{Full author list and affiliation list are presented at the end of the paper.}

\abstract[ABSTRACT]{We report the discovery of variable $\gamma$-rays up to petaelectronvolt (PeV; $\rm 1 \, PeV = 10^{15} \, eV$) from Cygnus X-3, an iconic X-ray binary. The $\gamma$-ray signal was detected with a statistical significance of approximately $10\sigma$ by the Large High Altitude Air Shower Observatory (LHAASO). Its intrinsic spectral energy distribution, extending from 0.06 PeV to 3.7 PeV, shows a pronounced rise toward 1 PeV after accounting for $\gamma$–$\gamma$ absorption by the cosmic microwave background radiation.    
We find variability on month-long timescales at a significance of $8.6 \sigma$, coinciding with a high state of the GeV gamma-ray flux detected by the Fermi-LAT. This,together with a $3.2\sigma$ evidence for orbital modulation, suggests that the PeV $\gamma$-rays originate within, or in close proximity to, the binary system itself. The observed energy spectrum and temporal modulation can be naturally explained  by $\gamma$-ray production through photomeson processes in the innermost region of the relativistic jet, where protons are accelerated to tens of PeV energies.}


\keywords{LHAASO, Cygnus X-3}

\maketitle

\section{Introduction}\label{sec1}

Cygnus X-3 is one of the first discovered X-ray binaries\cite{Giacconi:ApJL:1967}. 
It is a unique high-mass X-ray binary that comprises a Wolf–Rayet (WR) donor star\cite{vanKerkwijkNatur1992}, and a compact object, a black hole or a neutron star. The system has an exceptionally short period of 4.8~hours, 
and the orbital modulation has been detected in radio, infrared radiation, X-rays, and GeV $\gamma$-rays\cite{Becklin:Natur:1973,Molnar:Natur:1984,vanderKlis:A&A:1989,Tavani:Natur:2009,FermiLATCollaboration:Sci:2009,Zdziarski:MNRAS:2018}. 

Cygnus X-3 exhibits complex variability, occasionally producing major flares that reach peak radio fluxes of up to $\sim$20 Jy \cite{1995AJ....110..290W} and GeV $\gamma$-ray fluxes exceeding 
$10^{-6}$ ph cm$^{-2}$ s$^{-1}$ \cite{FermiLATCollaboration:Sci:2009}.
A classification of radio and X-ray states has been proposed based on the correlations between the X-ray flux and radio flux, which can generally be attributed to activities in the accretion disc and/or jet \cite{Szostek:MNRAS:2008,Tudose:MNRAS:2010,Koljonen:MNRAS:2010}.
GeV $\gamma$-ray flares predominantly occur during the soft X-ray state \cite{Tavani:Natur:2009,FermiLATCollaboration:Sci:2009,Koljonen:MNRAS:2010,Zdziarski:MNRAS:2018}.
The discovery by Very-Long-Baseline Interferometry (VLBI) of relativistic jets during radio flares established Cygnus X-3 as a 
microquasar\cite{Cygx3-jet1,Cygx3-jet2,Miller-Jones:ApJ:2004}. 
More recently, observations with the Imaging X-ray Polarimetry Explorer (IXPE), have provided new insight into the system’s accretion–ejection geometry. 
The detection of ~20\% X-ray polarization, oriented orthogonally to the radio jet axis, is interpreted as evidence for a collimated outflow, or a X-ray funnel\cite{CygX3-IXPE}. 
The intrinsic X-ray source is found to be obscured, while the derived apparent X-ray luminosity, if viewed face-on, would exceed $5 \times 10^{39}~{\rm erg}~{\rm s}^{-1}$, suggesting that Cygnus X-3 is a hidden ultraluminous X-ray (ULX) source\cite{CygX3-IXPE}.
Combined VLBI and IXPE results indicate that the radio jet propagates along and within the funnel structure\cite{Yang:MNRAS:2023}. 
A substantial fraction, from 10\% to nearly 100\% of the accretion power, may be channeled into the jet's kinetic energy\cite{CygX3-IXPE}, making Cygnus X-3 one of the most powerful known microquasars, with the outflow's mechanical luminosity approaching $10^{39}~{\rm erg}~{\rm s}^{-1}$.
Thus, “an astronomical puzzle named Cygnus X-3”\cite{CygX3-Hjellming}, after decades of intensive multi-wavelength studies, revealed itself as a representative of three extreme astrophysical source classes: microquasars, super-Eddington binaries, and ULX sources. 
Here, we claim yet another remarkable feature of this peculiar object - its operation as an extreme accelerator capable of accelerating protons beyond PeV, establishing Cygnus X-3 as a {\it Super PeVatron} \cite{Wang:ApJL:2025}.

Any discussion of Cygnus X-3 inevitably revisits a longstanding mystery from the early history of ground-based $\gamma$-ray astronomy. In the 1980s, the source attracted considerable attention following multiple claims of periodic TeV and PeV $\gamma$-ray signals. However, subsequent critical analyses\cite{CygX3-oldreview}, together with null results from more sensitive next-generation instruments, cast serious doubt on these early reports. One should mention, however, alternative explanations in the early judgments, relating the controversy to the variability (or episodic character) of the gamma-ray emission; see, e.g., \cite{Protheroe1994,Aharonian:SSRv:1996}.

Although the interpretation of these historical observations remains controversial and unclear, the community’s view on the plausibility of ultra-high-energy (UHE: $\geq0.1$~PeV) phenomena in accreting binary systems has evolved substantially. The shift has been driven by recent detections of $\gamma$-ray emission  extending into the UHE regime associated with
several prominent microquasars, including SS 433\cite{SS433-HAWC,SS433-HESS,SS433-LHAASO}, V4641 Sgr\cite{V4641-HAWC,SS433-LHAASO}, GRS 1915+105, MAXI J1820+070, and likely Cygnus X-1\cite{SS433-LHAASO}.
In this paper, we report the detection of a variable UHE $\gamma$-ray emission up to several PeV from Cygnus X-3, establishing it as yet another UHE microquasar, but distinct from all previously known UHE $\gamma$-ray sources due to its unique spectral and temporal properties. This detection was made possible by the exceptional performance of LHAASO.

LHAASO is a major extensive air shower (EAS) facility designed to study cosmic rays and gamma rays from TeV to PeV energies. Located at an altitude of 4.4 km above sea level in Sichuan Province, China\cite{he18}, LHAASO comprises three detector components: the Water Cherenkov Detector Array (WCDA), the Kilometer Square Array (KM2A), and the Wide Field-of-View Cherenkov Telescope Array (WFCTA) (see Ref.\cite{he18} for details). 
KM2A, designed for UHE gamma-ray observations, is composed of surface and underground detectors to register the electromagnetic and muon components of Extensive Air Showers (EAS) at energies from 10~TeV to 10~PeV. The combined data from these two subsystems enable effective suppression of cosmic-ray–induced background by selecting “muon-poor” EAS. Above 0.1 PeV, the cosmic-ray rejection power of KM2A exceeds $1 \times 10^4$, while retaining more than 80\% of $\gamma$-ray events\cite{km2a_fullarray_2024}. This capability, together with the gigantic detection area of  1.3~$\rm km^2$ and excellent angular resolution, ranging from  $0.24^{\circ}$ at  0.1~PeV to 0.1$^\circ$ at 1~PeV \cite{km2a_fullarray_2024,2021ChPhC..45b5002A}, provides a performance approaching an impressive sensitivity level of $10^{-14}~{\rm erg}~{\rm cm}^{-2}~{\rm s}^{-1}$ after several years of operation. 
Furthermore, KM2A allows effective searches for flux variability above 0.1 PeV on month-long timescales 
at a flux level as low as $10^{-12}~{\rm erg}~{\rm cm}^{-2}~{\rm s}^{-1}$.

\section{LHAASO Observations}\label{sec2}

\begin{figure*}[htbp]
\centering
    \includegraphics[width=1.0\textwidth]{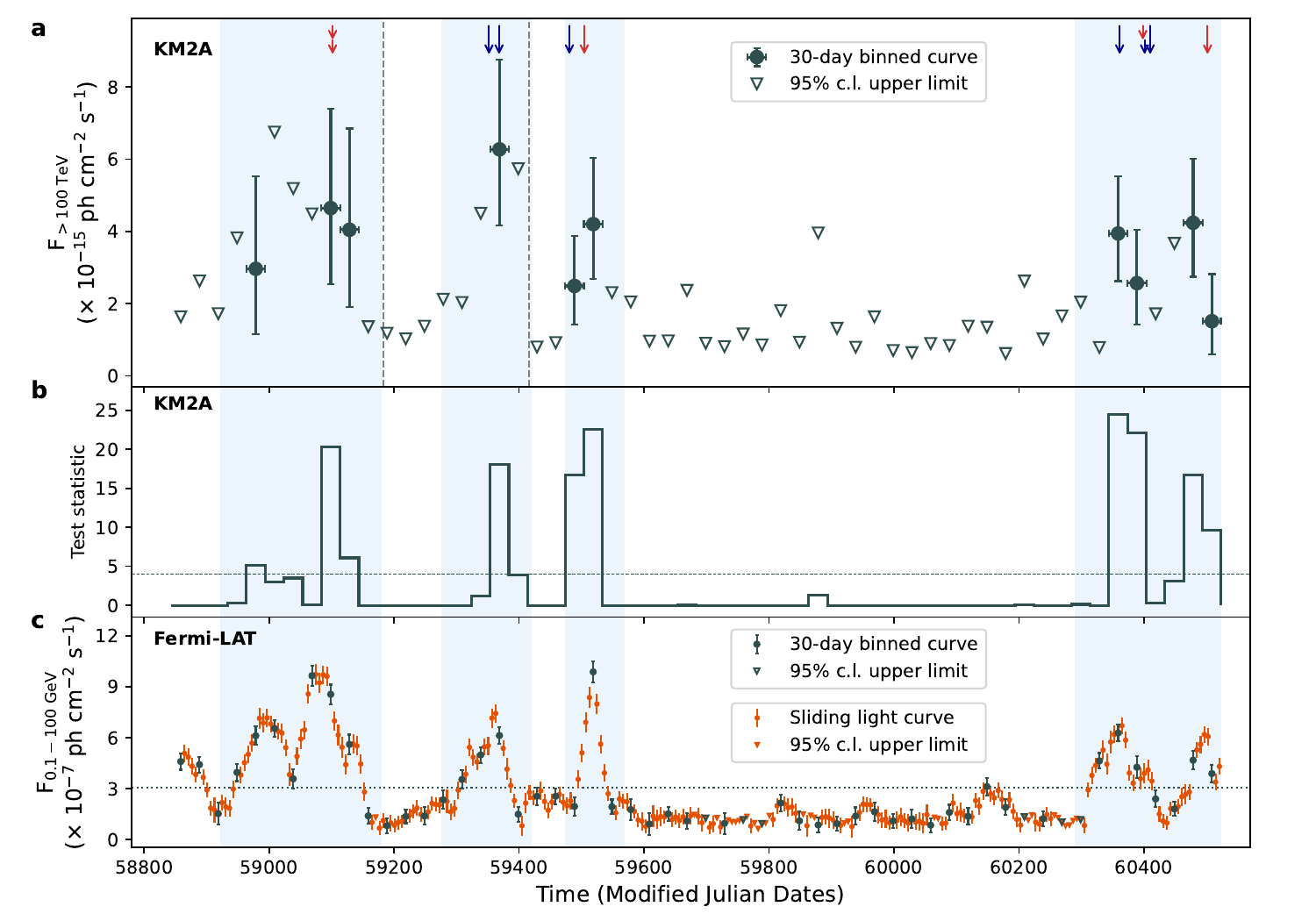}
    \caption{Fluxes and Test Statistic (TS) values of $\geq 0.1 \, \rm PeV$ $\gamma$-rays  from Cygnus X-3 as a function of time. 
    {\bf a}, 
 Flux above 0.1~PeV. The arrival times of individual high-energy photons in the 
 0.4–1~PeV and $\geq 1 \, \rm PeV$ ranges are indicated by blue and red arrows, respectively. 
 The observation began with half of the KM2A array in operation.
 Vertical dashed lines mark the commencement of operations for the three-quarter and full KM2A array configurations. {\bf b}, TS values corresponding to the detected $\geq$0.1 PeV $\gamma$-ray signals. For data with TS$\leq$ 4 (those below the horizontal dashed line), the 95$\%$ confidence-level flux upper limits are shown in the top panel.
{\bf c}, 
The 0.1–100~GeV light curve observed by {\it Fermi}-LAT. The horizontal dotted line indicates the average $\gamma$-ray flux. Time intervals where the sliding-window flux (orange points) exceeds this average define the $\gamma$-ray high states, shaded in light blue. The remaining periods are considered $\gamma$-ray low states. The first two flux points above the mean, part of a previous high-flux episode not fully covered by LHAASO, are excluded from the high-flux state classification.  
}
    \label{fig:light-curve}
\end{figure*}


The data analyzed in this work were collected with KM2A between December 2019 and July 2024, spanning three phases of the array.
Using KM2A data, we found a point-like source in the direction of Cygnus X-3, with a statistical significance of $ 9.6\sigma$. Detailed analysis procedures are described in Methods A. Remarkably, a significant signal is detected only at energies above approximately 100 TeV.
The signal is not uniformly distributed over time, as shown in Fig. 1, which presents the measured $\gamma$-ray fluxes (panel a) and corresponding Test Statistic (TS) values (panel b) for events above 0.1 PeV. 
For comparison, panel c shows the 0.1–100 GeV $\gamma$-ray light curve derived from the statistically rich {\it Fermi}-LAT data
\footnote{The light curve is obtained from the analysis of public {\it Fermi}-LAT data at \url{https://fermi.gsfc.nasa.gov/ssc/data/analysis/}. For details, see Methods.}.
Based on this, we defined high-state intervals, marked by the shaded pink zones in Fig. \ref{fig:light-curve}.
A visual correlation between the PeV and GeV gamma-ray fluxes is apparent in Fig. 1. 

To estimate the significance of the PeV signal during the $\gamma$-ray high and low states, accounting for the 1:1.6 ratio in exposure times between these two states, we modeled the 
$\gamma$-ray background using a template-based method described in Methods A. 
When restricting the analysis to the GeV-defined $\gamma$-ray high-state time window, the PeV $\gamma$-ray signal’s significance is increased to $ 11.5\sigma$, while it is below $2 \sigma$ during the $\gamma$-ray low state. The significance for variability of flux between the $\gamma$-ray high and low states is 8.6$\sigma$.

\begin{figure*}[htbp]
    \centering
    \includegraphics[width=1.0\textwidth]{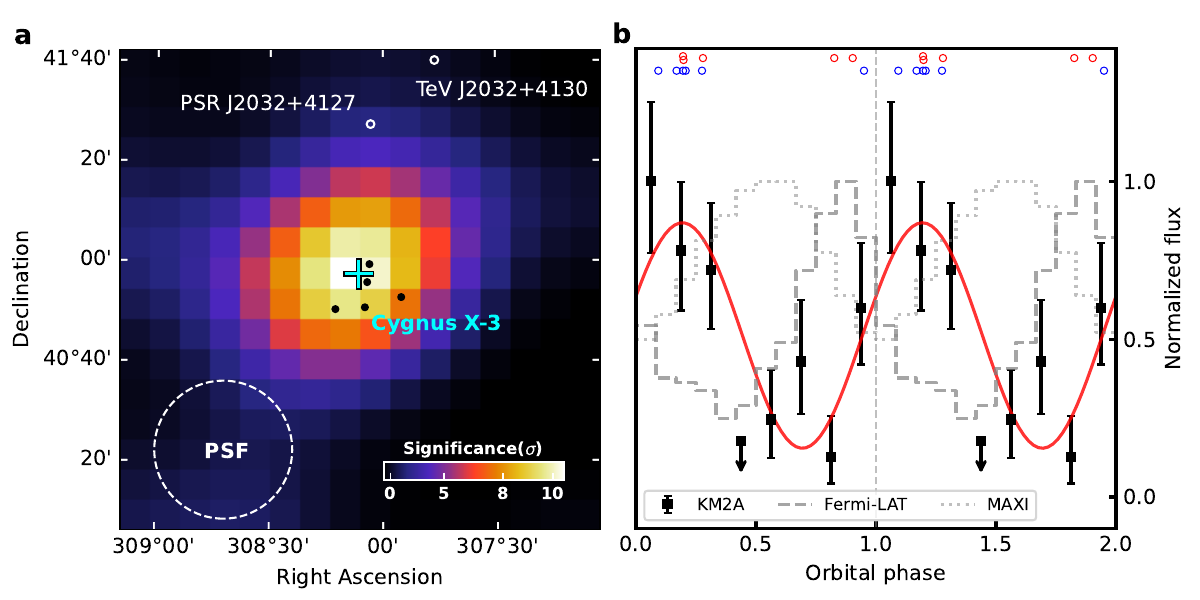}
    \caption{
{\bf a}, The significance map of $\geq 0.1$~PeV photons toward Cygnus X-3, based on KM2A observations during the $\gamma$-ray high state. The cyan cross marks the position of Cygnus X-3. Five $\geq 1$~PeV photons are shown as black dots. Contributions from background sources (indicated by open circles) have been subtracted. The white dashed circle indicates the 68\% point-spread function at 0.1~PeV.
{\bf b}, Orbital light curves of Cygnus X-3 
measured by LHAASO ($\geq$ 0.1~PeV), {\it Fermi}-LAT (0.1--100 GeV) and MAXI (2--20 keV).
For clarity, the fluxes are normalized using the following factors: $1.053 \times 10^{-15}$ 
for KM2A, $8.47 \times 10^{-7}$ for {\it Fermi}-LAT, and 1.36 for MAXI, all in units of ${\rm cm}^{-2}~{\rm s}^{-1}$. The red solid line represents a sinusoidal fit to the KM2A data, capturing the orbital modulation. The orbital phases of individual high-energy photons are marked at the top: $\geq 1$~PeV (red circles) and 0.4–1~PeV (blue circles).}
    \label{fig:skymap}
\end{figure*}

The significance map for $\geq 0.1$~PeV photons
detected during the gamma-ray high states of the source is shown in Fig.~\ref{fig:skymap} (panel a). The UHE emitter appears point-like, with best-fit coordinates of $\alpha$(J2000)=$308.10^{\circ}\pm0.03^{\circ}_{\rm stat}\pm0.03^{\circ}_{\rm syst}$ and $\delta$(J2000)=$40.92^{\circ}\pm0.02^{\circ}_{\rm stat}\pm0.03^{\circ}_{\rm syst}$,
consistent with the position of Cygnus X-3 (J2000 coordinates: $\alpha = 308.1074^{\circ}$, $\delta = 40.9578^{\circ}$ \cite{2004ApJ...600..368M}). 
In the same panel, we show that the other two nearby TeV sources are located approximately $0.5^\circ$ from Cygnus X-3, exceeding KM2A’s angular resolution of $0.24^\circ$ at 0.1 PeV.
Moreover, these two sources exhibit steep spectra above several tens of TeV (Method A).
Therefore, the contamination by these two sources is highly reduced above 0.1 PeV. This conclusion is further supported by the spatial clustering of five PeV photons within 10 arcminutes of Cygnus X-3 (see Fig. \ref{fig:skymap} and Table 1). 
Notably, two of these events, with reconstructed energies of $E = 3.73 \pm 0.41$~PeV and $3.08 \pm 0.34$~PeV, represent the highest-energy photons ever detected from an astrophysical source.

Adopting a distance of $9$~kpc\cite{Reid:ApJ:2023},
the observed upper limit for angular extension constrains the physical size of the $\gamma$-ray emitter to be no more than 44 light-years. 
The consideration of variability imposes tighter limits on the source physical size.
The arrival times of six photons with energies between 0.4 and 1 PeV and five photons above 1~PeV, are marked by blue and red arrows, respectively, in Fig.~\ref{fig:light-curve}. 
Notably, no photons with energies $\geq \, \rm 0.4 \, \rm PeV$ were detected during the $\gamma$-ray low state. 
The observed month-scale variability ($\Delta t\approx3$~months) of the gamma-ray emission suggests a sporadic origin, likely linked to episodic jet activity.
This temporal variability imposes a causality-based upper limit on the size of the emission region, constraining it to be within $R \sim c\Delta t \leq 2.3\times10^{17} \, \rm cm$. 
At this spatial scale, the inner jet of Cygnus X-3 is by far the most plausible site for particle acceleration and efficient gamma-ray production, although alternative scenarios cannot be entirely ruled out.


At GeV energies, the modulation of the $\gamma$-ray signal with the orbital period of Cygnus X-3 has already been observed with high statistical significance during the source’s high state by the {\it Fermi}-LAT collaborations\cite{FermiLATCollaboration:Sci:2009}. 
Remarkably, the UHE gamma-ray data alone show evidence of orbital modulation (see Fig.~\ref{fig:skymap}, right panel). The folded light curve as a function of orbital phase exhibits a clear peak near phase 0.2 and a minimum around phase 0.6. A likelihood analysis comparing the modulated signal to a constant-flux (null) hypothesis yields a post-trial statistical significance of 
$3.2 \sigma$.  
While this does not yet constitute definitive detection of orbital modulation of the PeV $\gamma$-ray signal, its correlation with the long-term (monthly) variability observed in the GeV band enhances confidence in the result. 
Taken together, these observations strongly support the conclusion that the PeV $\gamma$-ray emission originates on binary scales. 
As will be shown below, this is also supported by the PeV spectral hardening.

\begin{figure*}[htbp]
    \centering
    \includegraphics[width=0.8\textwidth]{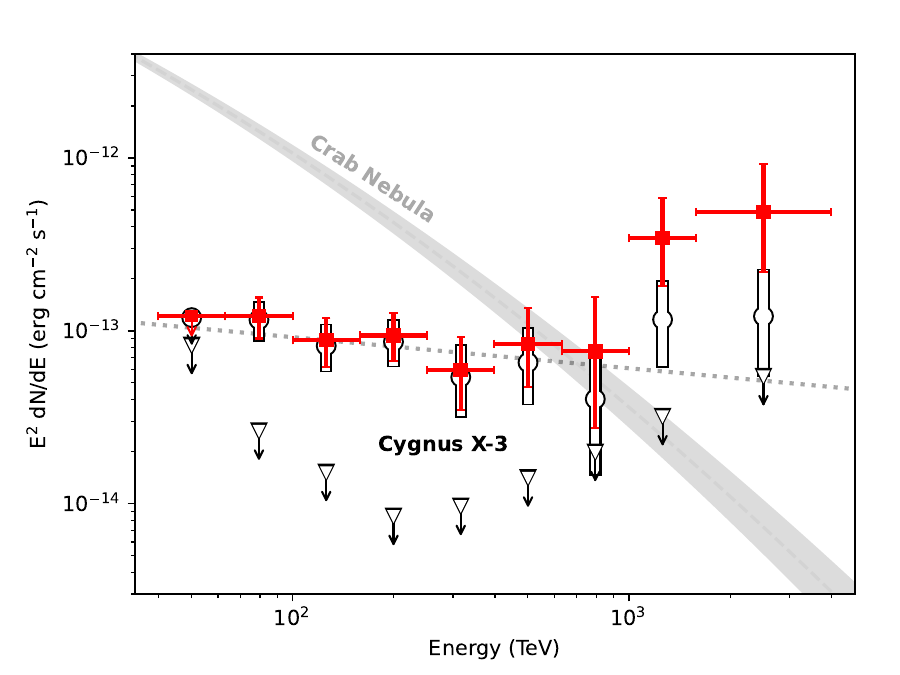}
    \caption{
UHE $\gamma$-ray SEDs. Open black circles represent fluxes measured in the high state. A power law fitting is presented as the grey dotted line. Red squares indicate the same fluxes but corrected for absorption by the ISRF and the 2.7~K CMB (Methods A.4). Open black triangles represents the flux upper limits in the $\gamma$-ray quiescent state. For comparison, the best-fit SED of the Crab Nebula, derived from KM2A data, is shown as a grey dashed band.}
    \label{fig:SED-LHAASO-obs}
\end{figure*}

The spectral energy distribution (SED) based on cumulative KM2A data collected during the high state of Cygnus X-3 is shown in Fig.~\ref{fig:SED-LHAASO-obs}. 
For comparison, we also include the best-fit spectrum of the Crab Nebula\cite{km2a_fullarray_2024}, a well-established point-like UHE $\gamma$-ray source extending to 1 PeV. The spectrum of Cygnus X-3 is peculiar and differs from those of all previously reported TeV/PeV sources.
A strongly significant signal is detected above 0.1 PeV, with detection significance of 
$7.1 \sigma$, $4.6 \sigma$, and $6.4 \sigma$  in the 0.1 - 0.4~PeV, 0.4 - 1~PeV, and $>$1~PeV intervals, respectively. 
The signal below 0.1 PeV reaches only $3.5\sigma$, primarily contributed by a narrow energy interval between 63 and 100 TeV. No significant excess is detected in the WCDA data at energies below 20~TeV. 
In Fig.~\ref{fig:SED-LHAASO-obs}, we also present flux upper limits for the $\gamma$-ray low state, which reveal nearly an order of magnitude suppression of the UHE $\gamma$-ray emission compared to the high state.

We fit the measured flux values shown in Fig.~\ref{fig:SED-LHAASO-obs} with a power-law spectrum $dN/dE=N_{0}(E/E_{0})^{-\Gamma}$ and reference energy $E_{0}$ chosen to be 50 TeV. 
The flux normalization is $N_{0}$=(2.6$\pm$0.6$)\times10^{-17}~$TeV$^{-1}~$cm$^{-2}~$s$^{-1}$ and the photon index is $\Gamma=2.18 \pm 0.14$, characterizing Cygnus X-3 the hardest UHE source ever detected by LHAASO \cite{LHAASO-1st-catalog}. 
There is a possible spectral hardening above 1 PeV, and this trend becomes more pronounced when accounting for the absorption of UHE $\gamma$-rays during their propagation from the source to the observer.

 At TeV–PeV energies, $\gamma$-ray absorption is primarily due to $\gamma\gamma$ pair production ($\gamma + \gamma \rightarrow e^+ + e^-$) on low-energy photon fields. In general, we find that absorption of TeV–PeV photons by the infrared–millimeter (IR–mm) and X-ray radiation fields of the binary system is negligible. Absorption by the ultraviolet (UV) photon field mainly affects TeV energies and becomes rather  weak above $\sim$100 TeV (see Methods C.3).

Additional absorption occurs as TeV–PeV photons propagate through the interstellar medium, where they interact with the Cosmic Microwave Background (CMB) and the interstellar radiation field (ISRF), which includes starlight (optical) and dust (infrared) 
emission components \cite{Popescu:MNRAS:2017}. 
For PeV photons, absorption is dominated by the CMB. 
For the known distance to the source, this permits a high-precision calculation of the optical depth $\tau(E_\gamma)$. Accordingly, the intrinsic gamma-ray spectrum, $ J_0 (E) = J_{\rm obs}(E) \exp[\tau (E)]$, can be robustly reconstructed. Notably, photons in the 2-3 PeV range experience the strongest absorption, by a factor of up to 3, while the effect decreases at both lower and higher energies (see Methods A.4). 
The correction amplifies the trend of spectral hardening above 1~PeV.
This feature is seen in Fig. \ref{fig:SED-LHAASO-obs}, where the absorption-corrected fluxes are shown alongside the observed fluxes.

\section{Hadronic origin of the UHE emission}\label{sec3}
Taking the distance of $d_{X-3} \approx 9$~kpc, 
the UHE $\gamma$-ray luminosity of Cygnus X-3 is $L_{\rm \gamma, UHE}\approx 10^{33}~{\rm erg}~{\rm s}^{-1}$, which is only a small fraction of its intrinsic X-ray luminosity, estimated\cite{CygX3-IXPE} as high as $10^{39}~{\rm erg}~{\rm s}^{-1}$. 
This intrinsic luminosity is comparable to
the Eddington luminosity of a black hole of mass $M_{\rm BH} \leq 10 M_\odot$. 
In the high state, a substantial fraction of the system's power can be channeled into the kinetic power of the inner jet. This region appears to be the most realistic, if not the only plausible, site at the binary scale where efficient particle acceleration can occur. 
As the accelerator’s characteristic size is limited by the dimensions of the compact system
($R \sim 3\times 10^{11} \, \rm cm$; e.g., \cite{Zdziarski:MNRAS:2018,Antokhin:2021qhp}), electrons and protons can reach PeV energies only in the presence of a strong magnetic field, following the Hillas criterion (see Method C.1),
$B \geq  11 \, (E/{\rm 1 \, PeV})  \,  (R/3\times10^{11} \, \rm cm)^{-1}\beta^{-1} \, G$, where $\beta=v/c$ is the jet velocity in the acceleration region.

On the other hand, when both particle acceleration and synchrotron cooling are taken into account, the maximum electron energy is given by $E_{e, \rm max}= 0.06 \, \eta^{1/2}  
(B/1 \, \rm G)^{-1/2} \, \rm PeV$ (see Method C.1).
Combining the constraints from synchrotron cooling and the Hillas criterion yields an upper limit on the electron energy:
 $E_{e, \rm max} \leq  0.07 \, \eta^{1/3} \beta^{1/3} (R/3\times10^{11} \, \rm cm)^{1/3} \, \rm PeV$, which robustly excludes leptonic origin of observed 
 PeV radiation in any realistic scenario. This limitation, however, does not apply to the periodic emission observed at lower (GeV) energies, which is best explained by anisotropic Compton scattering of 1–100 GeV electrons accelerated in the inner jet\cite{Dubus:MNRAS:2010,Zdziarski:MNRAS:2012,Zdziarski:MNRAS:2018,Dmytriiev:ApJ:2024}.
For protons, synchrotron cooling is significantly less restrictive than for electrons. In magnetic fields on the order of 100~G, protons can be accelerated to energies of tens of PeV, as permitted by the Hillas condition. While such strong magnetic fields may arise in the inner jets of powerful microquasars, in the case of Cygnus X-3, the field strength cannot exceed $\sim$1~kiloGauss in the emission region; otherwise, PeV $\gamma$-rays would be absorbed by the magnetic field via electron–positron pair production. This constraint implies that proton acceleration in Cygnus X-3 is limited to energies below $\sim$100~PeV.


Within the binary system, TeV and PeV $\gamma$ rays are generated through hadronic  $pp$ (e.g., \cite{Hillas1984,Aharonian:SSRv:1996,Romero:A&A:2003}) and $p\gamma$ \cite{Aharonian1985,Levinson:PhRvL:2001} processes --  via the production and subsequent decay of $\pi^0$ mesons.
The photohadronic ($p \gamma$) channel is particularly compelling due to the high density of stellar photons, which offsets the relatively low cross-section of these interactions \cite{Aharonian1985}. The combination of the Wolf–Rayet star’s intense ultraviolet radiation with a luminosity as high as $10^6 L_\odot$ in  $\sim 10 \, \rm eV$ photons, and the extreme compactness of the system ($R \approx 3 \times 10^{11}$~cm), 
renders Cygnus X-3 exceptional among known X-ray binaries. Notably, the "neutrality" of photons makes them an ideal target for photomeson production, as they do not disrupt the jet structure. In contrast, hadronic interactions with dense gaseous environments, such as the stellar wind, could significantly impact the jet propagation \cite{Perucho:A&A:2010,Bosch-Ramon:A&A:2016,Lopez-Miralles:A&A:2022}.


Photomeson production in $p \gamma$ interactions has a strict kinematic threshold, given by
$2 E_{p} \epsilon (1-\beta_p \cos \theta) \geq (2 m_p m_\pi+ m_\pi^2) c^4$ (see e.g. Ref.\cite{Kelner:PhRvD:2008}).  For relativistic protons ($\beta_p \to 1$), this condition is reduced  to $E_{p} \geq 140 \, (1-\cos \theta)^{-1} \epsilon_{\rm eV}^{-1}   \, \rm PeV$, where 
$\epsilon_{\rm eV}=\epsilon/1 \, {\rm eV}$ is the target photon energy, and $\theta$ is the collision angle. 
Taking the average energy of the starlight photons in Cygnus X-3, $\epsilon \approx 10$~eV, and assuming a proton-photon interaction angle $\theta = 90^\circ$, the energy of proton should exceed $\approx 10 \, \rm PeV$ to initiate photomeson production. Given that approximately 10 \%  of the proton energy is transferred to the resulting $\gamma$-rays \cite{Kelner:PhRvD:2008}, one naturally expects a sharp increase in the SED above 1 PeV. 

Due to the resonance peak in the inelastic $p\gamma$ cross-section at 200–500 MeV in the proton rest frame, the $\gamma$-ray flux is expected to decline at higher energies, especially when the target photon distribution is narrow (Planckian).  Even for a hard power-law spectrum of protons, this leads to a drop in the $\gamma$-radiation spectrum.
 Additionally, a high-energy cutoff in the proton distribution would 
further sharpen the spectral break. 
At low energies, the $\gamma$-ray spectrum also drops off due to the kinematic threshold of the interaction.

\begin{figure*}[htbp]
    \centering
    \includegraphics[width=0.9\textwidth]{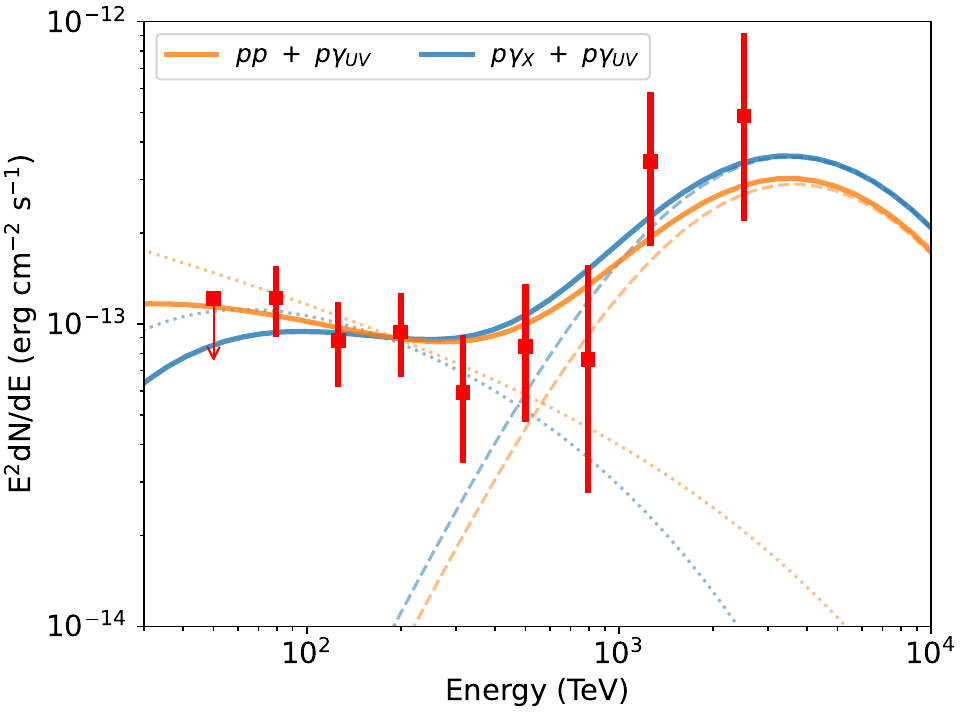}
    \caption{
Broadband modelling of the SED of Cygnus X-3 with two scenarios: $pp+p\gamma_{UV}$ and $p\gamma_{X}+p\gamma_{UV}$ (see Methods \ref{Methods:Theory} for details). 
Solid lines are the sum of individual components with the correction of $\gamma\gamma$ absorption by the companion UV photon field.
The contributions of individual components without correction for $\gamma\gamma$ absorption are shown with corresponding dotted and dashed lines.}

    \label{fig:SED-LHAASO-model}
\end{figure*}

Below 1~PeV, the suppression of the $\gamma$-ray flux from proton-starlight interactions is too steep to account for the measured fluxes down to around 0.06~PeV. Therefore, an additional radiation channel is needed to explain the measured fluxes in this energy range.  Under certain conditions, $p\gamma$ interactions between lower-energy (TeV-PeV) protons and X-ray photons, together with $pp$ interactions within the jet, and potentially also outside the jet (e.g., with the stellar wind), may contribute significantly and help bridge the spectral gap between 10~TeV and 1~PeV (Methods C).
In scenarios involving $pp$ interaction, a number density of $\sim2\times10^{10}~{\rm cm}^{-3}$ 
can account for the observed SED as shown in Fig.~\ref{fig:SED-LHAASO-model}. 
This is close to the gas number density in the jet (Methods C.2).



As noted, the extension of the $\gamma$-ray spectrum to lower energies could be explained by an additional photomeson production channel involving interactions between multi-TeV protons and X-ray photons within the jet. The compact object in Cygnus X-3 exhibits an intrinsic X-ray luminosity of $\sim 10^{39}~{\rm erg}~{\rm s}^{-1}$, comparable to the UV luminosity of its companion Wolf-Rayet star. While the X-ray photon production rate is substantially lower (by a factor of $\epsilon_{\rm X}/\epsilon_{\rm UV} \sim 10^2-10^3$), the local X-ray photon density can be significantly enhanced if particle acceleration occurs preferentially near the compact object rather than the companion star.
This scenario naturally emerges if protons are accelerated and isotropized near the jet base, where they are exposed to intense X-ray emission from either the accretion disc or the jet funnel. The corresponding $\gamma$-ray spectrum predicted by this model is shown in  Fig.~\ref{fig:SED-LHAASO-model}. 

The kinematic threshold for $p \gamma$ interactions introduces a strong angular dependence in the $\gamma$-ray production rate (Fig. M8), giving preference to specific orientations of the jet and resulting in an effective modulation of PeV radiation. This interpretation is supported by the concentration of {\it all} detected photons with energies exceeding 0.4~PeV within two narrow orbital phase intervals, 0.1–0.3 and 0.8–1.0 (Fig. \ref{fig:skymap}, right panel), which likely correspond to the most favorable interaction geometries.
The PeV flux modulation has a conceptual similarity to the 4.8-hour orbital modulation observed at GeV energies, as both signals originate from interactions with the UV photon field of the companion star.
At GeV energies, modulation can arise from anisotropic inverse Compton (IC) scattering, where the cross section depends on the angle between interacting photons and relativistic electrons. 
At PeV energies, the modulation is driven by the angular dependence of the energy threshold for $p \gamma_{UV}$ interactions.
While both phenomena reflect the system’s geometry, the resulting light curves may be governed by distinct conditions and timescales. 
Furthermore, GeV and PeV $\gamma$ rays likely originate from different locations
(see Methods C.1 for a discussion).
Consequently, the relation between the modulated GeV and PeV light curves could be quite complex without a necessity for a correlation. 
At lower energies, modulation of $pp$ and $p\gamma_X$ interactions may result from the complex interplay between outflows driven by the compact object (i.e., the jet and wind) and the WR wind (see Methods C.2 for a discussion). In addition, at $\sim$100~TeV, mild orbital modulation may arise due to anisotropic, albeit weak, absorption by the UV photon field.

\section{Conclusion and Discussion}\label{sec4}

In conclusion, the detection of variable UHE $\gamma$-ray emission up to 3.7 PeV from Cygnus X-3, characterized by a possible 4.8-hour modulation and a distinctive spectral pileup/bump above 1 PeV, suggests efficient proton acceleration — likely within the inner jet — to energies exceeding 15 PeV (Methods C.4). 
The spectral pileup (or bump) can be naturally interpreted as arising from photomeson interactions between multi-PeV protons and UV photons from the Wolf–Rayet companion star. The emission below 1 PeV is likely produced via hadronic interactions with either X-ray photons from the accretion inflow/outflow or plasma inside the jet and the stellar wind. Neutrinos are expected to be generated in these interactions at comparable flux level.
While ultra-high-energy $\gamma$-ray halos recently discovered around several microquasars suggest indirect links to their central engines, this observation provides the first compelling evidence that a microquasar can host a super-PeVatron and generate transient PeV $\gamma$-ray emission in close proximity to the binary system. 

Finally, we note that PeV protons accelerated in the jet, upon escaping into the interstellar medium, will unavoidably produce an extended halo of UHE $\gamma$ rays around Cygnus X-3. Depending on the combination of ambient medium density and diffusion coefficient, an average proton injection power significantly $\gtrsim~10^{37}~{\rm erg}~{\rm s}^{-1}$ could render this halo detectable by LHAASO. Such a halo provides a plausible alternative explanation for the Cygnus Bubble \cite{CygnusBubble_2023}, suggesting that Cygnus X-3 (rather than the Cygnus OB2 stellar cluster) may be the primary engine powering this UHE $\gamma$-ray structure, despite its significantly greater distance from the observer (see Methods C.5).

\section*{Acknowledgements}
We would like to thank all staff members who work at the LHAASO site above 4400 meters above sea level year round to maintain the detector and keep the water recycling system, electricity power supply and other components of the experiment operating smoothly. We are grateful to the Chengdu Management Committee of Tianfu New Area for the constant financial support for research with LHAASO data. We appreciate the computing and data service support provided by the National
High Energy Physics Data Center for the data analysis in this paper. 
This work is based on public Fermi-LAT data provided by the NASA Goddard Space Flight Center and on the Fermitools provided by the Fermi Science Support Center.
This research work is supported by the following grants: The National Natural Science Foundation
of China (NSFC) No.12522510, No.12393851, No.12393852, No.12393853, No.12393854, No.12205314, No.12105301, No.12305120, No.12261160362, No.12105294, No.U1931201, No.12375107, NO.12273038, CAS Project for Young Scientists in Basic Research(No. YSBR-061),Youth Innovation Promotion Association
CAS (No.2022010, No.2023275) and in Thailand by the National Science and Technology Development Agency (NSTDA) and the National
Research Council of Thailand (NRCT) under the High-Potential Research Team Grant Program (N42A650868).

\section*{Author Contributions}
Cong Li proposed this project and analysed the data. Zhen Cao coordinated this specific working group. 
Jieshuang Wang performed the theoretical calculations and the modelling. 
Jieshuang Wang and Felix Aharonian led the interpretation of the results. 
Jianeng Zhou was responsible for multi-wavelength data analysis and temporal analysis. 
Felix Aharonian, Zhen Cao, Cong Li, Jieshuang Wang and Jianeng Zhou prepared the manuscript. Shicong Hu and Dmitri Semikoz cross-checked the result. The whole LHAASO collaboration contributed to the publication, with involvement at various stages ranging from the design, construction and operation of the instrument to the development and maintenance of all software for data calibration, data reconstruction and data analysis. All authors reviewed, discussed and commented on the present results and on the manuscript.

\bibliographystyle{nsr}
\bibliography{ref}

\newpage
\onecolumn
\begin{center}
    \noindent
Zhen Cao$^{1,2,3}$,
F. Aharonian$^{3,4,5,6}$,
Y.X. Bai$^{1,3}$,
Y.W. Bao$^{7}$,
D. Bastieri$^{8}$,
X.J. Bi$^{1,2,3}$,
Y.J. Bi$^{1,3}$,
W. Bian$^{7}$,
A.V. Bukevich$^{9}$,
C.M. Cai$^{10}$
% ...
(The LHAASO Collaboration),
and 
J.S. Wang$^{7,6,37}$

\noindent
$^{1}$ Key Laboratory of Particle Astrophysics \& Experimental Physics Division \& Computing Center, Institute of High Energy Physics, Chinese Academy of Sciences, 100049 Beijing, China\\
$^{2}$ University of Chinese Academy of Sciences, 100049 Beijing, China\\
$^{3}$ TIANFU Cosmic Ray Research Center, Chengdu, Sichuan,  China\\
$^{4}$ University of Science and Technology of China, 230026 Hefei, Anhui, China\\
$^{5}$ Yerevan State University, 1 Alek Manukyan Street, Yerevan 0025, Armeni a\\
$^{6}$ Max-Planck-Institut for Nuclear Physics, P.O. Box 103980, 69029  Heidelberg, Germany\\
$^{7}$ Tsung-Dao Lee Institute \& School of Physics and Astronomy, Shanghai Jiao Tong University, 200240 Shanghai, China\\
$^{8}$ Center for Astrophysics, Guangzhou University, 510006 Guangzhou, Guangdong, China\\
$^{9}$ Institute for Nuclear Research of Russian Academy of Sciences, 117312 Moscow, Russia\\
$^{10}$ School of Physical Science and Technology \&  School of Information Science and Technology, Southwest Jiaotong University, 610031 Chengdu, Sichuan, China\\
% ...
$^{37}$ Max Planck Institute for Plasma Physics, Boltzmannstra{\ss}e 2, D-85748 Garching, Germany
\end{center}



\section*{Supplementary material}

# Methods

# A    Data analysis

## A.1    Detector and its performance:

The Large High Altitude Air Shower Observatory (LHAASO) comprises three detector systems: the 1.3 km$^2$ Array (KM2A), the Water Cherenkov Detector Array (WCDA), and the Wide-Field Cherenkov Telescope Array (WFCTA). This study utilizes data from KM2A, the primary LHAASO component for detecting gamma rays with energies above 25 TeV. KM2A consists of two sub-arrays: a surface array, which measures the electromagnetic component of air showers (electrons and gamma rays), and an underground array, which measures the muon content of the showers. A detailed description of the LHAASO detectors is provided in Ref. (*22*).

The performance of LHAASO-KM2A, calculated by averaging the zenith angle distribution along the trajectory of Cygnus X-3 on the sky, is shown in Fig. M1. The left panel shows the Point Spread Function (PSF), defined as the angle $\phi_{68}$ that contains 68 % of the events (left X-axis). It is approximately 0.24° at 0.1 PeV, reduced to 0.1° above 1 PeV. The parameter $\rho_{50}$, defined as the particle density at 50 m from the shower axis and derived by fitting the modified NKG lateral distribution function (*24*), is used to calculate the energy of the primary photon. The energy resolution is improved from 17 % at 0.1 PeV to 10 % at 1 PeV. The ratio between the measured numbers of muons and electrons in a shower is used to discriminate electromagnetic showers from hadronic showers. The rejection power is about $1.0 \times 10^4$ at energies around 100 TeV . The details about the calibration and shower reconstruction and selection are described in Ref. (*23, 24*). Fig. M2 shows the differential sensitivity for sources with different times of exposure, demonstrating the potential for LHAASO-KM2A to detect transient signals with different time scales.

## A.2    Data analysis:

The 3-dimensional ('3-D') fitting framework is used to separate signals from individual sources (*23*). The source list and templates developed in Ref. (*33*) are used to model gamma-ray background for Cygnus X-3. The previous study revealed a huge UHE gamma bubble extending to at least 6° at

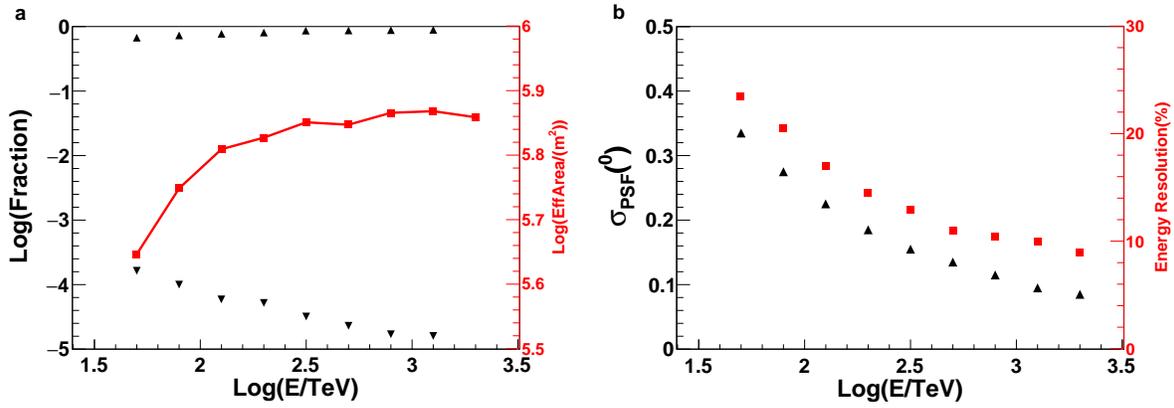

**Figure M1**: The performance of LHAASO-KM2A.**a**: Survival fraction of $\gamma$-rays(black) and cosmic ray background(blue) after the discrimination cuts. The effective area, averaged by the zenith angle distribution of Cygnus X-3, as a function of energy is shown as the red line. **b**: Angular resolution(black) and energy resolution(red) at different energy bins.

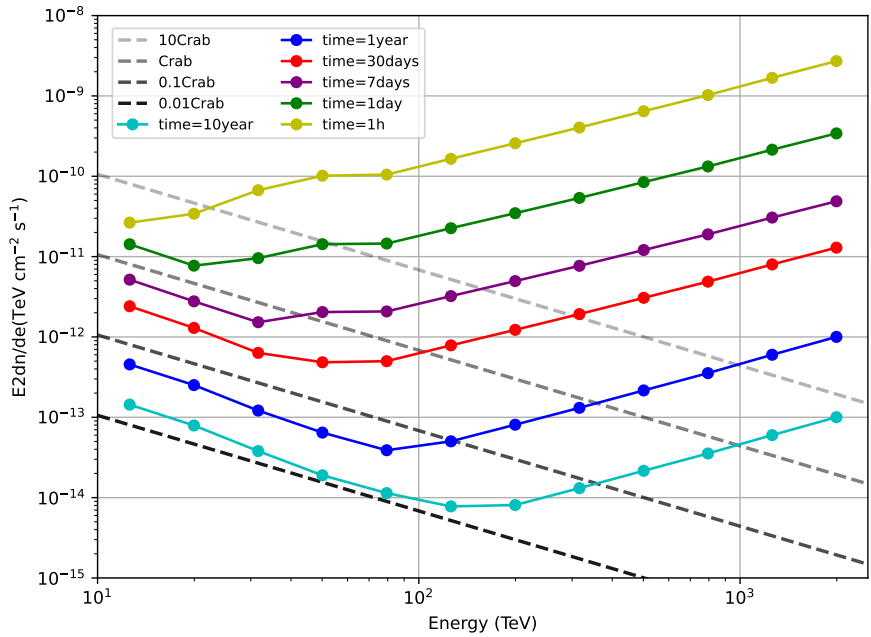

**Figure M2**: The differential sensitivity of LHAASO-KM2A with different observational times.

the Cygnus region, named as the Cygnus bubble. Cygnus X-3 is located at the center of the Cygnus Bubble. The Cygnus bubble is modeled by a combination of a Gaussian distribution, $H_2$ distribution and the HI distribution. Besides the Cygnus bubble, there are other UHE gamma-ray sources located within the Cygnus region. Among them, LHAASO J2031+4157u is the nearest bright source, which is at an angular distance of about $0.51°$ from Cygnus X-3. LHAASO J2031+4157u was separated into two sources in a previous study, e.g., LHAASO J2031+4141 and LHAASO J2032+4125 (*34*). The SED of LHAASO J2032+4125 is best-fitted by an exponential cutoff function $dN/dE = N_0(E/E_0)^{-\Gamma} \exp^{-E/E_{cut}}$. The estimated parameters are $N_0 = (1.5 \pm 0.7) \times 10^{-15}$ TeV$^{-1}$ cm$^{-2}$ s$^{-1}$, $\Gamma = 1.2 \pm 0.5$ and $E_{cut} = 30 \pm 8$ TeV. The spectrum of LHAASO J2031+4141 is consistent with a very steep (with $N_0 = (3.9 \pm 1.6) \times 10^{-17}$ TeV$^{-1}$ cm$^{-2}$ s$^{-1}$ and $\Gamma = 4.0 \pm 0.3$) power law function. We added a new source in the fitting iteration until the improvement of TS is less than 25. We found that the new source is a point-like source with best-fit position $\alpha(J2000) = 308.11 \pm 0.03_{stat} \pm 0.03_{sys}$, $\delta(J2000) = 40.90 \pm 0.03_{stat} \pm 0.03_{sys}$, which is consistent with the position of Cygnus X-3. The spectrum is fitted by a power-law function with an index of $2.01 \pm 0.10$, which is the hardest spectrum ever detected by LHAASO-KM2A.

We binned the data into time intervals of 30 days to obtain the light-curve of Cygnus X-3. The binned likelihood fitting is performed at each time interval. The array live time is calculated for each time bin. To ensure the convergence of the fitting, all the parameters for background sources are fixed during the fitting. Only the normalization for Cygnus X-3 is left free. Considering there is only one parameter, the square root of TS is the significance. The integral flux is calculated by integrating above 100 TeV using the normalization obtained by fitting. For comparison, the light curve of the Crab nebula is also derived using the same method to test the stability of detectors at the same time. As shown in Fig.M3, the flux is consistent with a constant flux within the statistical errors, which verifies the stable operational status of the detector.

To further explore the correlation between the TeV and GeV light curves, the data are divided into two parts according to GeV light curve. Considering there is no indication of variability for background sources, the flux for background sources are pre-assumed to be constant with time. Besides background sources, a point source is added during GeV flaring or quiescent times. The TS is increased by 148.4 by adding a new point source at flaring times, corresponding a significance of 11.5 $\sigma$. The position of the new source is consistent with that Cygnus X-3. For comparison, the

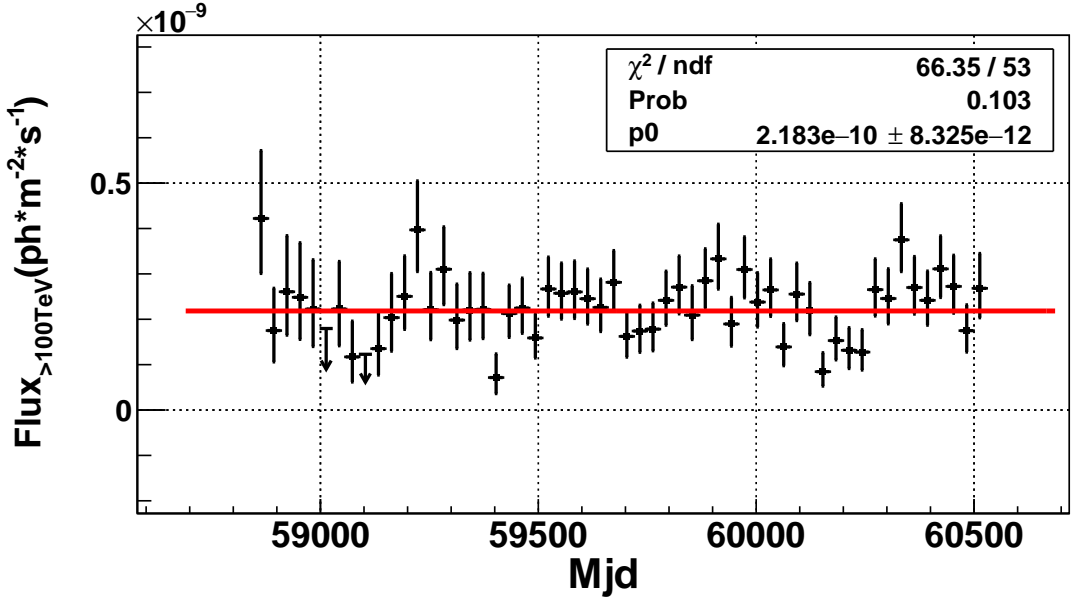

**Figure M3**: The light curve of the Crab nebula above 100TeV, which is consistent with a constant flux, verifying the stability of KM2A.

increase of TS is only 3.8 at quiescent times. The spectral index during active state is $\Gamma = 2.18\pm0.14$. The $\Delta$TS between hypotheses of variable TeV fluxes at two GeV flux states and constant flux all the time is 75, resulting a significance of $8.6\sigma$. The residual significance map after subtraction of contributions from all background sources is shown in Fig.M4. No significant excess was detected during quiescent times, which not only demonstrates the sporadic nature of the signal but also verifies the reliability of background model.

The data is binned according to the orbital phase of the source to derive the orbital light curve. A similar likelihood fitting process is performed at each phase bin to obtain the integral flux. The $\Delta$TS between constant flux assumption and variable flux at each phase is 28.24 with 8 bins, corresponding to a significance of $3.5\sigma$. The post-trial significance is estimated by assuming 3 trials (three different binnings were tried), corresponding a significance of $3.2\sigma$. We also use a sine function to fit the points, and the improvement of TS is 15.24. Considering there are two free parameters added, the significance is $3.3\sigma$.

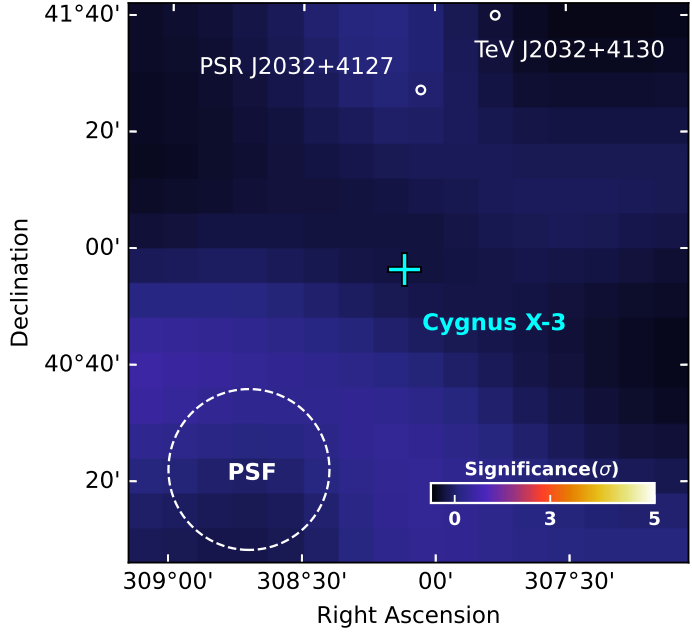

**Figure M4**: Residual significance map with energy ≥ 0.1 PeV after subtraction of contributions from all background sources during the quiescent state.

## A.3 The highest energy photons:

Benefiting from the excellent rejection power of LHAASO, it can almost achieve a background free observation for a point source above 50 TeV. The detailed information of events with energy above 0.4 PeV within 95% angular range is listed in Table.1. The angular range is chosen according to the detector PSF and pointing accuracy. The cumulative probability of $\log_{10}(N_\mu/N_e)$ with cosmic rays, which are recorded with similar $\Theta$ and $dr$ but higher energy of photon-like events, are calculated using experimental data. The total number of cosmic rays from the Cygnus X-3 direction during the flaring state is used to obtain the probability of misidentifying cosmic ray events as a photon-like event.

## A.4 Correction for the ISRF and CMB absorption:

Cygnus X-3 is located in the Galactic plane at a distance of $8.95 \pm 0.96$ kpc (*25*). The gamma rays are subjected to absorption due to the $\gamma\gamma$ interaction with the photons of the ISRF and CMB. Here we adopt the ISRF from Ref. (*26*). The optical depth for their absorption is shown in Fig. M5.

| E (TeV) | $\delta$E (TeV) | $N_e$ | $N_\mu$ | $\Theta(°)$ | $dr$(m) | $\phi(°)$ | $P_{CR}(> E)$ | ToA (MJD) |
|---------|-----------------|-------|---------|-------------|---------|-----------|---------------|-----------|
| 1476 | ±133 | 6355 | 23.7 | 11.61 | 20.8 | 0.16 | $5.7 \times 10^{-3}$ | 59101.609 |
| 1421 | ±128 | 6258 | 6.6 | 12.73 | 57.6 | 0.04 | $1.5 \times 10^{-4}$ | 59101.625 |
| 512 | ±46 | 3736 | 4.0 | 25.66 | 75.3 | 0.12 | $6.1 \times 10^{-3}$ | 59351.843 |
| 518 | ±52 | 2374 | 3.6 | 32.38 | 88.2 | 0.21 | $1.1 \times 10^{-2}$ | 59367.984 |
| 578 | ±107 | 2984 | 6.5 | 21.36 | 106.8 | 0.16 | $1.3 \times 10^{-2}$ | 59480.632 |
| 1188 | ±107 | 5480 | 14.1 | 34.41 | 71.9 | 0.10 | $3.7 \times 10^{-3}$ | 59504.394 |
| 483 | ±44 | 2212 | 8.9 | 24.33 | 42.5 | 0.11 | $2.5 \times 10^{-1}$ | 60361.085 |
| 3735 | ±411 | 21926 | 72.4 | 31.41 | 139.7 | 0.04 | $2.3 \times 10^{-4}$ | 60398.160 |
| 929 | ±84 | 4459 | 8.7 | 23.78 | 59.8 | 0.13 | $2.1 \times 10^{-3}$ | 60400.980 |
| 805 | ±73 | 3737 | 3.3 | 11.97 | 61.2 | 0.21 | $3.9 \times 10^{-4}$ | 60410.015 |
| 3086 | ±340 | 19567 | 70.4 | 16.13 | 118.2 | 0.12 | $3.5 \times 10^{-4}$ | 60501.812 |

**Table 1**: Specifications of the PeV photons. E and $\delta$E are the reconstructed energy and its error. $N_e$ and $N_\mu$ are the detected numbers of secondary charged particles and muons. $\Theta$ is the incident zenith angle of the shower. $dr$ is the core distance from the edge of LHAASO-KM2A. $\phi$ is the angular distance between the event and Cygnus X-3. $P_{CR}(> E)$ is the probability of misidentifying a cosmic-ray event as a photon-like event. ToA is the time of arrival of each event in MJD.

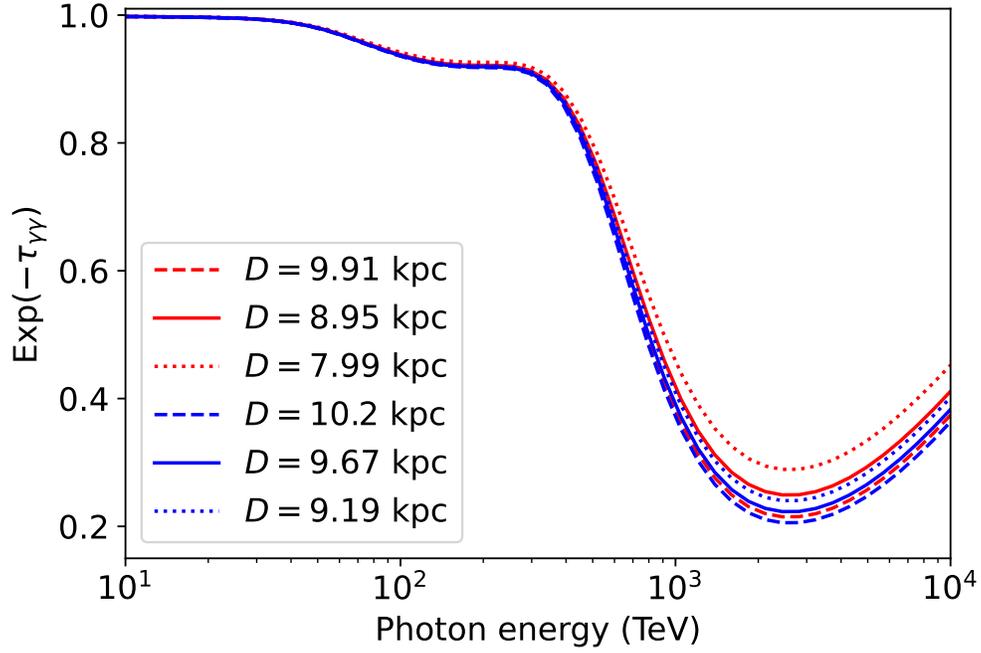

**Figure M5**: The absorption factor of the $\gamma\gamma$ interaction with the ISRF and CMB for a source at a distance of $D = 8.95 \pm 0.96$ kpc or $D = 9.67^{+0.53}_{-0.48}$ kpc (*25*).

The ISRF dominates the optical depth over $10 - 200$ TeV, and it is mild with $\tau \sim 0.1$, therefore, different choices of the ISRF model will not affect our conclusion significantly. At higher energies, $> 200$ TeV, the absorption by the CMB takes over. The optical depth increases significantly at photon energies $200$ TeV $\lesssim$ E $\lesssim 2$ PeV. The optical depth peaks at $\sim 2$ PeV with $\tau \sim 1$, and then decreases as a power-law.

## A.5 Systematic uncertainties

The systematic errors affecting SED were previously investigated by studying Crab Nebula. The biggest contribution is from the deviation between the real atmosphere density profile and atmosphere model used in simulation due to seasonal and daily changes. The uncertainties are estimated to be 7% for the flux and 0.02 for index. The likelihood method,taking advantage of the shape of the point spread function(PSF) in fitting, is used in this analysis to separate the signals from individual sources, which may introduce uncertainty from the PSF. The PSFs from both simulation and Crab Nebula data are used in fitting to estimate this uncertainty, which results in a change of

flux of about 12% and index of 0.04. Thus, the total uncertainties for the flux and index are 14% and 0.045, respectively.

# B Multi-wavelength data analysis

## B.1 Fermi-LAT:

For analysis of the GeV *Fermi*-LAT data, we select the photon events with energy between 0.1 and 100 GeV that are spatially located within a 10° radius region around the position of Cygnus X-3. The Pass 8 CLEAN event class (corresponding instrument response function `P8R3_CLEAN_V3`) is used, benefiting from its lower background rate. Sources from the Fermi's fourth source catalogue (*34, 35*), galactic diffuse emission (model `gll_iem_v07`) and extra-galactic isotropic emission (model `iso_P8R3_CLEAN_V3_v1`) are used to model the backgrounds. Since the catalogue description does not cover the time of our analysis, we performed binned likelihood analysis within the time range of MJD 58848.0 and MJD 60522.0 as a global fitting to derive the best-fit parameters for further analysis. Although the GeV spectrum of Cygnus X-3 is described by a log-parabola model in the catalogue (*34, 35*), here we still tried a power-law model fitting for the data not included in the catalogue for comparison. The TS statistics, calculated by comparing their likelihood values, favors the log-parabola model with TS = $-2\Delta\log(\text{likelihood}) \approx 112$.

The light curve is binned into 30-day intervals, which is consistent with the binning of KM2A data. Binned likelihood analysis is performed to obtain the gamma-ray flux in each bin. For the fitting convergence in this complicated region, except for Cygnus X-3 and variable sources within 5°, we fix the spectral shapes of all background sources based on the previous best-fit model, only with varying normalization. The GeV light curve is shown in Fig. 1. To determine the gamma-ray active periods, we used the average flux as the threshold. Here the sliding-bin method was used, since it can catch more details in the light curve (*36, 37*). In practice, a 30-day bin slides with a step of 6-day is used over the entire time span and we individually calculate the gamma-ray flux at each bins. In such a "smoothed" light curve, we can clearly see five distinguished flares above the average flux level. We choose the nearest minimal-flux times (below the averaged flux) as the starting and ending time of a given period. The last two periods under such criteria are connected, so they are

combined into one. Finally, the four high-state periods are determined as MJD 58921.7-59179.7, MJD 59275.7-59419.7, MJD 59473.7-59569.7, and MJD 60289.7-60523.0, respectively, as marked in Fig. 1 with light-pink areas. It is worth noting that although the first two points are above the average flux, they are part of a previous high-state period not covered by LHAASO operation, so we did not include them into analysis. Furthermore, data after July 31, 2024 (MJD 60523.0) are not included since there is a cutoff for LHAASO data selection, although the GeV flare has not ended at this time.

We obtained the orbital light curve by calculating fluxes over 12 independent phase bins with data during high-state periods. The arrival times of all photons were corrected for barycentering, and then orbital phases were assigned based on the parabolic ephemeris (*27*), where the zero phase was defined from the X-ray minimum time (*38*). In each orbital phase bin, we reselected the events based on their assigned phases and performed likelihood estimation of the flux of Cygnus X-3. Notably, we did not apply pulsar gating to PSR J2032+4127 as in Ref. (*6*) due to the unavailability of the latest pulsar ephemeris. Instead, we analyzed all Cygnus X-3 high-state data directly, a treatment also used by Ref. (*27*) and proven to have no significant impact on the Cygnus X-3 analysis. Our resulting profile aligns with the most recent results (*27*) but slightly differs from Ref. (*6*), which might stem from differences of early event reconstruction of LAT or intrinsic changes in Cygnus X-3.

## B.2  MAXI:

The MAXI X-ray monitoring of Cygnus X-3 reveals its variability in the 2–20 keV band [1]. Here, we used one-orbit binned light curve (binsize=1.5h) to construct the orbital variability. Similar to the correction for GeV data, the times were first corrected for barycentering, and then histogrammed into an orbital light curve based on the parabolic ephemeris (*27*) and assigned phases. The resulting orbital light curve is shown in Fig. 2.

---

[1] http://maxi.riken.jp/star_data/J2032+409/J2032+409.html

## B.3 The correlation between GeV and TeV-PeV light curves:

We employed the pyZDCF (*39*) package to compute the correlation between the TeV-PeV and GeV light curves. In practice, since the TeV-PeV light curve contains a large number of upper-limit measurements, which cannot be directly handled by ZDCF, we adopted an alternative strategy: the upper-limit points were assigned different weighting factors and then treated as normal data points when calculating the correlation with the GeV light curve. Specifically, we defined $f = w^{\xi} f_m$, where $f_m$ is the measured flux (either a normal data point or an upper limit) and $w$ is the weighting factor. Here, $\xi = 0$ for normal points and $\xi = 1$ for upper-limit points. We tested a range of weighting factors from 0.1 to 1.0 in steps of 0.1, to evaluate their impact on the correlation results (see inset of Fig. M6). Encouragingly, under all tested weights, the correlation peak between the TeV-PeV and GeV light curves consistently appears at lag $\simeq$0, indicating that the correlation is robust given the current temporal resolution of the light curves. To further quantify the significance of the correlation peak at lag $\simeq$0, we performed Monte Carlo simulations of $10^5$ TeV-PeV light curves and computed their ZDCFs with the GeV light curve. By examining the distribution of correlation coefficients at different lag values, we determined the confidence level of the observed correlation between TeV-PeV and GeV light curves. To generate the simulated TeV-PeV light curves, we adopted the `simulator` method in the `stingray` (*40*) package, which uses statistical properties such as the mean flux and rms variability extracted from the original TeV light curve. For each choice of weighting factor, we independently simulated $10^5$ light curves to compute the corresponding confidence level of the correlation. In Fig. M6, we present the correlation function for the case of $w = 0.5$, together with the $1\sigma$, $3\sigma$, and $5\sigma$ confidence levels (red dashed lines). The correlation peak clearly exceeds the $5\sigma$ level. Furthermore, the confidence levels of the peaks obtained under different weighting factors are shown in the inset. While the significance decreases as the weight increases, even for $w = 1.0$, the correlation peak still exceeds the $3.8\sigma$ level.

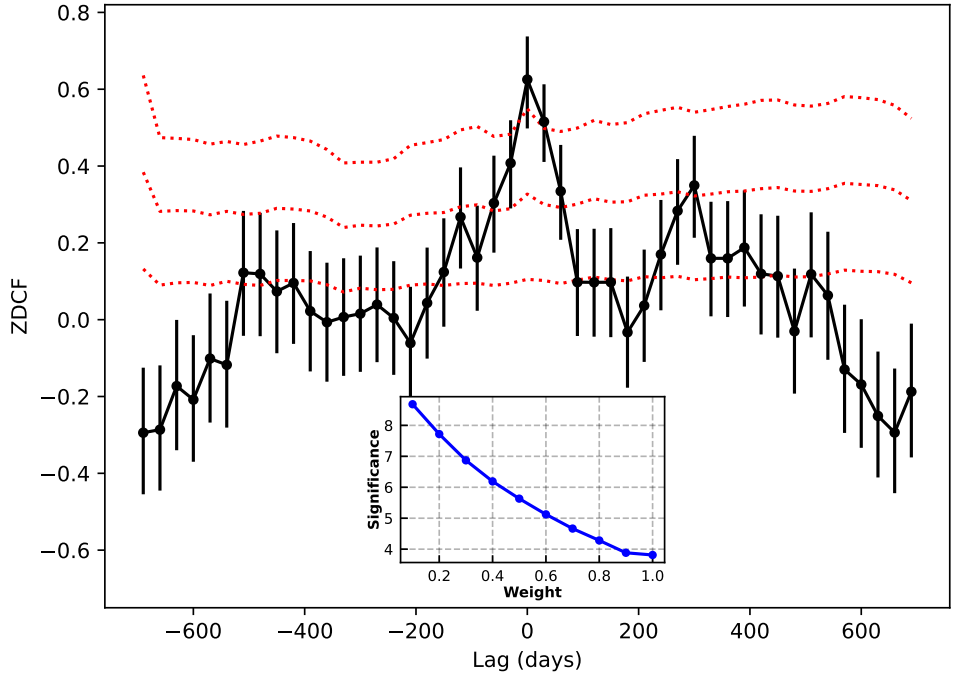

**Figure M6**: The Z-transformed discrete correlation function (ZDCF) between TeV and GeV light curves. The peak at lag≈0 indicates the strong correlation between them. Different weights are estimated for upper-limit data points when calculating correlations. Here is the ZDCF for weights 0.5, and a simulation from $10^5$ artificial light curves gives a significance of $5.6\sigma$ for the correlation. Red dotted lines from bottom to top are $1\sigma$, $3\sigma$ and $5\sigma$ confidence levels, respectively. The correlation significance under different weights is shown in the inset plot.

# C  The origin of UHE gamma rays

## C.1  General requirements for particle acceleration and emission site:

It can be found that to explain the gamma-ray SED, charged particles need to be accelerated to at least multiple PeV energies regardless of the emission process. We parameterize the acceleration time as $t_{acc} = \eta r_g/c$, where the particle mean free path is cast to be proportional to the gyro radius $r_g$. The maximum energy of particles is limited by the electric potential of the system, i.e., the Hillas criterion ($E_{max} = eBr_j\beta$). Meanwhile, particles needs to be accelerated within the dynamical time, which is $t_{acc} \leq z_j/\beta c$ taking a jet geometry as Fig. M7, where $z_j$ is the jet length. Combining both effect, we obtain

$$E_{max} = eBr_j\beta = 30B_3 r_{j,11}\beta \min[1, z_j/(\eta\beta^2 r_j)] \text{ PeV}, \tag{1}$$

where the system magnetic field and size are $B = 10^3 B_3$ G and $r_j = 10^{11} r_{j,11}$ cm, respectively. This would correspond to a Poynting flux of $L_B = \frac{B^2}{4\pi}\pi r_j^2\beta c \approx 0.75 \times 10^{38} B_3^2 r_{j,11}^2$ erg s$^{-1}$, which can be achieved for Cygnus X-3 considering its kinetic power can reach $L_K \sim 10^{39}$ erg s$^{-1}$. Therefore for efficient acceleration, protons can be accelerated to energies above 10 PeV in the jet of Cygnus X-3.

In such a high magnetic field, PeV gamma rays can be absorbed by the magnetic field and converted into electron-position pairs, under the condition $E_\gamma B_\perp/(m_e c^2 B_{cr}) \gtrsim 0.1$, where $B_\perp$ is the magnetic field perpendicular to the photon momentum, $B_{cr} = 4.14 \times 10^{13}$ is the quantum critical field, and $m_e$ is the electron mass (*41*). In panel b of Fig. M8, we show the optical depth of $\gamma B$ absorption for photons propagating in a magnetic field with $B_\perp = 10^3$ G to a distance of $R = 10^{11}$ cm. To make sure that absorption is insignificant for the detected $E_\gamma = 4$ PeV photon, the required magnetic field in the emission site is $B_\perp < 10^3$ G. Assuming $B \sim B_\perp$ and taking $\beta = 0.5$ (*12, 42*), the size of the emission region should be $r \gtrsim 10^{11}$ cm to allow $E_{max} \gtrsim 15$ PeV according to Eq. (1). We note that it is suggested that the magnetic field $B \lesssim 100$ G in the GeV emitting region (*31*). One possibility is that the GeV and TeV-PeV emitting regions are slighly different (e.g., different jet heights $z_j$), which is also supported by the possible phase shift in the orbital light curve (panel b of Figure 2).

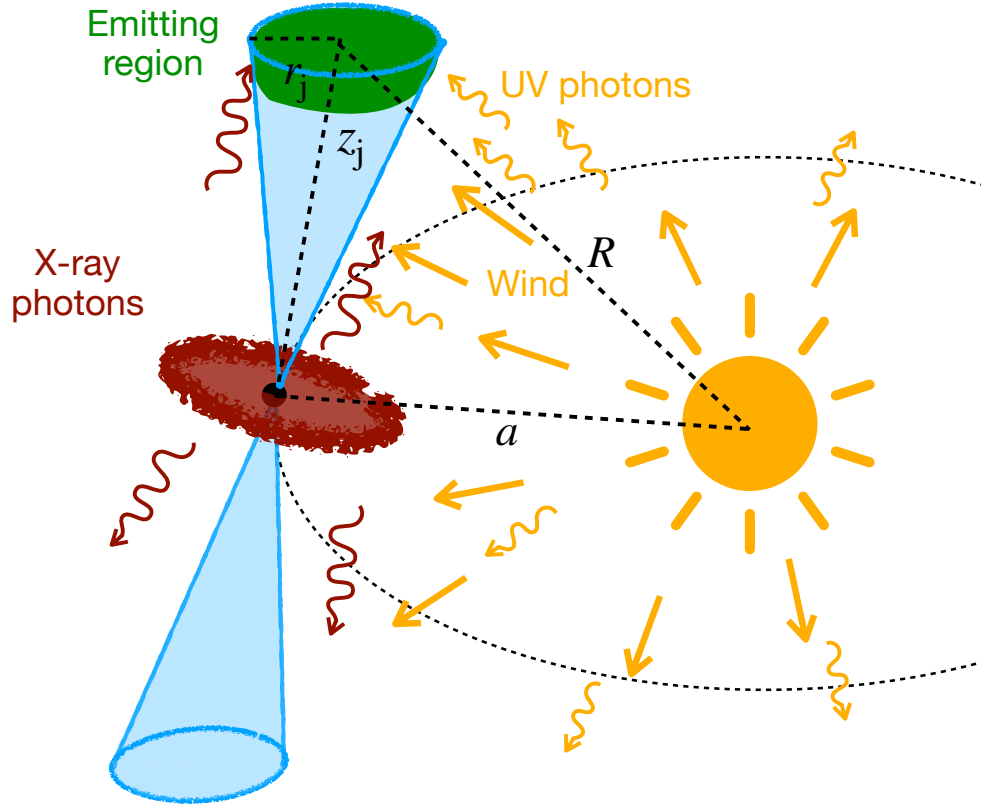

**Figure M7**: A schematic figure for the Cygnus X-3 system (not to scale). The emission region is separated by $z_j$ and $R$ from the compact object to companion star, respectively. The jet radius is denoted by $r_j$. The binary separation is taken to be $a = 2.65 \times 10^{11}$ cm (*27, 28*). The dense wind from the WR star can impact the jet propagation depending on its moment flux, which can bend or even disrupt the jet, e.g., Refs. (*43, 44*). X-rays are produced during the accretion process, which could originate from the inner disc, corona, outflow or reflection by the inner funnel (*45, 46*). Particles can be accelerated in the jet, and produce hadronic interactions with the X-ray and UV photons ($p\gamma$ interaction), or with the jet material ($pp$ interaction). Note that the X-ray funnel suggested by the X-ray polarisation measurement (*13*)is not shown in this cartoon. This cartoon represents the simplest scenario for this system, while the actual interaction between WR wind and the accretion inflow/outflow can be more complicated.

## C.2 Hadronic processes for the UHE emission:

It has been found that the GeV emission and its orbital modulation can be explained by the IC scattering between electrons from the inner jet and the anisotropic photon field from the companion star (*27, 30, 31*). For the UHE emission, both the indication of orbital modulation at UHE energies and the association with GeV flares suggest that it is also produced at the binary scale, close to the GeV emission site. However, the radiation processes are quite different for GeV and UHE gamma rays. As shown above, electron suffer from significant synchrotron cooling in the binary scale, and thus cannot reach PeV energy, so that only hadronic processes for UHE emissions are possible.

Cygnus X-3 contains a Wolf-Rayet (WR) star with a strong wind and UV emission. In Fig. M7 we show the basic geometry of the system. In such a system, both $p\gamma$ and $pp$ processes can occur in the jet, where the target photon field comprises UV photons from the companion and/or X-rays from the accretion process, and the target material is the jet material for the $pp$ interaction. In the panel a of Fig. M8, we show the interaction time for $p\gamma$ and $pp$ processes for different number densities, as well as the orbital time scale. As the emission is likely modulated by the orbital period, the gamma-ray production time should be smaller than the orbital period. This requires a target density $n_p \gtrsim 10^{11}$ cm$^{-3}$ for the $pp$ interaction, and $n_\gamma \gtrsim 3 \times 10^{13}$ cm$^{-3}$ for the $p\gamma$ interaction. Note that isotropic interactions are assumed here, while for anisotropic interactions, a factor of $(1 - \cos\theta)$ correction should be applied to the target photon energy, where $\theta$ is the interaction angle.

For the $pp$ interaction, the jet density can be derived from its kinetic power. With a four velocity of $\Gamma\beta c$, the jet kinetic luminosity is $L_K \approx (\Gamma - 1)n_j m_p c^2 \pi r_j^2 \beta c$, where $m_p$ is the proton mass. The corresponding number density is $n_j \approx 10^{10} L_{K,39} r_{j,11}^{-2}$ cm$^{-3}$ with $r_j = 10^{11} r_{j,11}$ cm and $L_K = L_{K,39} 10^{39}$ erg s$^{-1}$. Therefore, the jet is not dense enough unless the kinetic luminosity reaches $L_{K,39} \gtrsim 10$. In principle, the WR wind can be more dense that the jet, however, in this case it is unclear whether the emission would preserve orbital periodicity. Therefore, we focus on the $p\gamma$ interactions in the following discussion.

The UV photons come from the WR star. Following Ref. (*27*), we adopt a binary separation $a = 2.65 \times 10^{11}$ cm assuming a circular orbit, with the companion WR star producing blackbody emission at a temperature $T_* \sim 10^5$ K and a radius $R_* \sim 10^{11}$ cm. The starlight density from the companion is $n_{UV} \approx 7 \times 10^{14} R_{*,11}^2 T_{*,5}^3 (R/a)^{-2}$ cm$^{-3}$ at a separation of $R \approx a$, where the

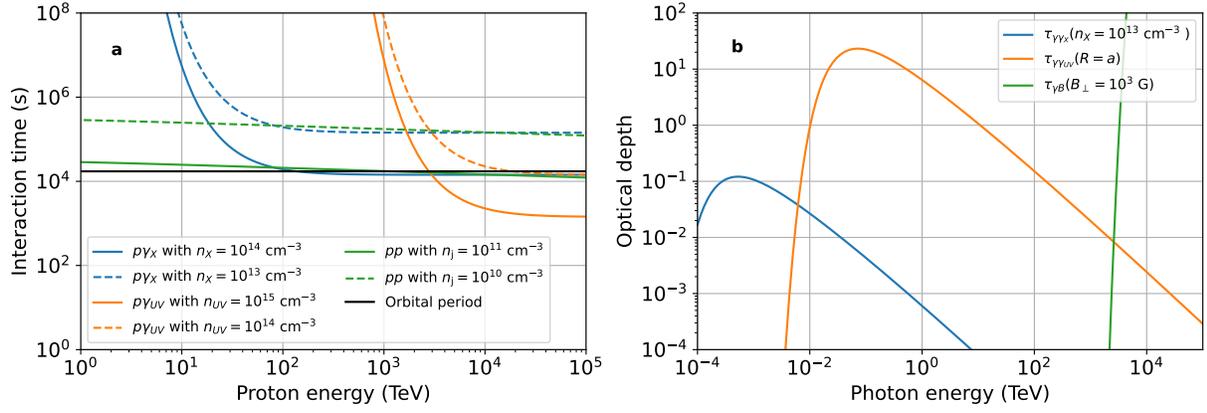

**Figure M8**: Panel a: The gamma-ray production time for $p\gamma$ and $pp$ interactions with different number densities in comparison with the Cygnus X-3 period. Panel b: The optical depths for different photon absorption processes.

WR-star radius is $R_* = R_{*,11}10^{11}$ cm and the temperature is $T_* = T_{*,5}10^5$ K. The X-rays could originate from the inner disc, corona, or the outflow (*45, 46*). During the gamma-ray high state, modeling of the thermal component of X-rays gives a temperature $T_{\mathrm{accretion}} = 1.4 \times 10^7$ K and a flux $8.1 \times 10^{-9}$ ergs cm$^{-2}$ s$^{-1}$. The X-ray spectrum is derived through MAXI ondemand (*47*), and fitted with `XSPEC` (v. 12.13.1). The corresponding number density is $n_X \approx 10^{13}z_{\mathrm{j},11}^{-2}$ cm$^{-3}$, where $z_{\mathrm{j}} = z_{\mathrm{j},11}10^{11}$ cm. However, it has been pointed out that the intrinsic X-ray flux can be more than ten times higher than the observed one, as indicated by the X-ray polarimetry observations (*13*).

Notably, as illustrated in panel a of Fig. M8, a spectral hardening feature at PeV energies is naturally expected due to the $p\gamma$ interaction with the photon field from the WR star. The $p\gamma$ interaction has a threshold energy at (*32*)

$$E_p \approx 14[E_t(1 - \cos\theta)/10\,\mathrm{eV}]^{-1}\,\mathrm{PeV}, \tag{2}$$

where $E_t$ is the target photon energy. The produced photon energy is $E_\gamma \sim 0.1E_p$. Therefore, for the UV photon field from the WR star, the produced photon energy is $E_\gamma \gtrsim 1/(1 - \cos\theta)$ PeV, which can naturally account for the spectral hardening at $\gtrsim 500$ TeV. For the X-ray photon field with $E_t \sim 1$ keV, the produced gamma rays have energy $E_\gamma \gtrsim 14$ TeV. In the following we consider $p\gamma$ interactions with both UV and X-ray photon field as an example for understanding the UHE spectrum.

## C.3 $\gamma\gamma$ absorption:

When hadronic interactions are important, gamma-ray absorption processes can also be effective. Gamma rays produced in the inner jet can suffer from $\gamma\gamma$ absorption by X-ray or UV photons or absorption by the magnetic field. The optical depth is $\tau_{\gamma\gamma} = \int n_{X/UV}\sigma_{\gamma\gamma}dl$, where the integration is along the line of the sight. As the X/UV photon density decreases significantly with distance to sources with $n_X(z_j) \propto z_j^{-2}$ or $n_{UV}(R) \propto R^{-2}$, we approximate the optical depth with $\tau_{\gamma\gamma X} \approx n_X\sigma_{\gamma\gamma}z_j$ and $\tau_{\gamma\gamma UV} \approx n_{UV}(R)\sigma_{\gamma\gamma}R$. This is appropriate for a mildly relativistic jet (*12, 42*). Adopting the isotropic absorption cross-section, the optical depths are presented in panel b of Fig. M8. It shows that in general the absorption can be important for gamma rays with $E_\gamma \lesssim 100$ TeV. Thus we take into account the absorption due to the UV photon field for the theoretical spectral modelling below.

## C.4 Examples of spectral modeling:

We adopt an analytical model (*32*) to calculate the secondary spectrum with anisotropic effects for $p\gamma$ interaction. We also ignore the Doppler effect in the modeling, because the correction should be insignificant as the jet is found to have a mild relativistic velocity. Given the large free parameter space for the spectral modelling, here we fix some parameters to provide examples. We assume a power-law spectrum of protons with a cutoff at $E_c = 50$ PeV to account for the UHE spectrum: $dN/dE \propto E^{-\alpha_p}\exp(-E/E_c)$. For the target X-ray photon field, we assume a density $n_X = 3 \times 10^{13}$ cm$^{-3}$ and an interaction angle $\theta_X = 69°$. As discussed above, the $p\gamma_{UV}$ process can naturally explain this spectral hardening; thus we include it for the same population of accelerated protons. For the companion stellar field, we adopt a temperature $T_* = 10^5$ K and a distance $R = a$, so the corresponding photon density at $R = a = 2.65 \times 10^{11}$ cm is $n_{UV} \approx 7.2 \times 10^{14}R_{*,11}^2$ cm$^{-3}$. Adopting a spectral index $\alpha_p = 2.3$, an interaction angle $\theta_{UV} = 79°$, and a total energy budget of protons $W_p(E > E_0) = 4.8(E_0/1 \text{ TeV})^{-0.3} \times 10^{39}$ erg, we obtain the SED shown in orange lines in Figure 4. We found that the intrinsic SED can be well reproduced. Assuming that the length of the emission zone is scaled to the jet height ($z_j$), the power of high-energy protons can be estimated as $L_p(E > E_0) = W_p(E > E_0)\beta c/z_j = 7 \times 10^{38}(E_0/1 \text{ TeV})^{-0.3}z_{j,11}^{-1}$ erg/s. This should be smaller than jet kinetic luminosity, i.e. $L_p < L_K$, requiring Cygnus X-3 to be a super-Eddington source. This is consistent with the X-ray polarimetry results, which indicates that Cygnus X-3 is a

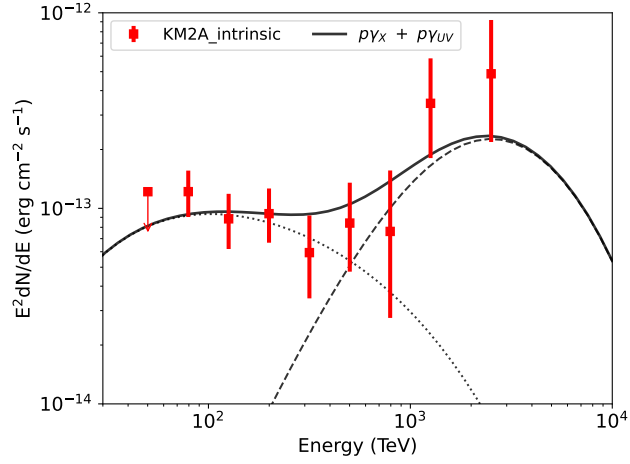

**Figure M9**: An example of SED modeling similar to Fig. 4, but with proton cutoff energy at 15 PeV.

hidden ULX (*13*). We also tried SED modeling assuming $E_c = 15$ PeV with $\alpha_p = 2.2$, $\theta_{UV} = 80°$, $\theta_X = 72°$, as shown in Fig. M9. At such a cutoff energy, the spectral peak of the $p\gamma_{UV}$ interaction is located at around 2 PeV. A lower cutoff energy would lead to a lower peak energy, deviating from observations. Thus, to explain the detected spectrum $E_c \gtrsim 10$ PeV is generally required, indicating that Cygnus X-3 is a proton super-PeVatron. In Fig. M10, we show that adopting the scattering angle for the $p\gamma_{UV}$ process ($\theta_{UV} = 79°$), the GeV SED can also be explained with an electron spectral index $\alpha_e = 3.4$, and a cutoff at 26 GeV based on the anisotropic IC scattering model (*27, 30, 31*). Such a cutoff energy is allowed even considering the strong cooling effect.

For the $p\gamma$ scenario, the periodic modulation may be explained by changes in the interaction angles over the orbital period. For $p\gamma_{UV}$, this would occur when the jet is orbiting with the companion, as suggested for the anisotropic IC scenario. For $p\gamma_X$, this might also occur when the jet direction is modulated by the stellar wind. The interaction angle can significantly modify the flux, as shown in Fig. M11. In this case, the interaction angles used in the above spectral modeling should be interpreted as 'average' angles over the orbital phase.

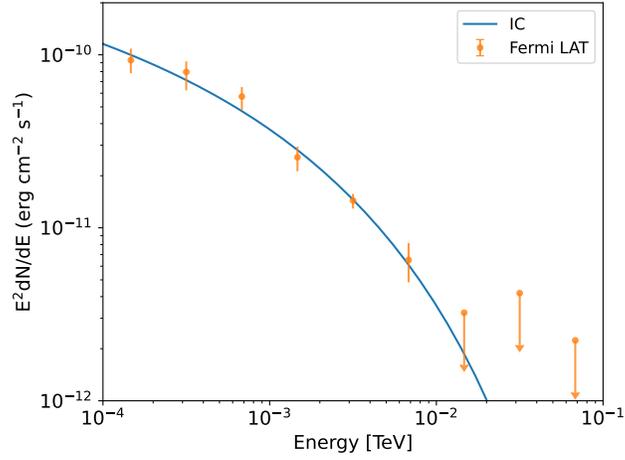

**Figure M10**: SED modeling of GeV data based on the anisotropic IC scattering model (*27, 30, 31*). Here we adopt the scattering angle $\theta_{UV} = 79°$, the same as the angle for the $p\gamma_{UV}$ process.

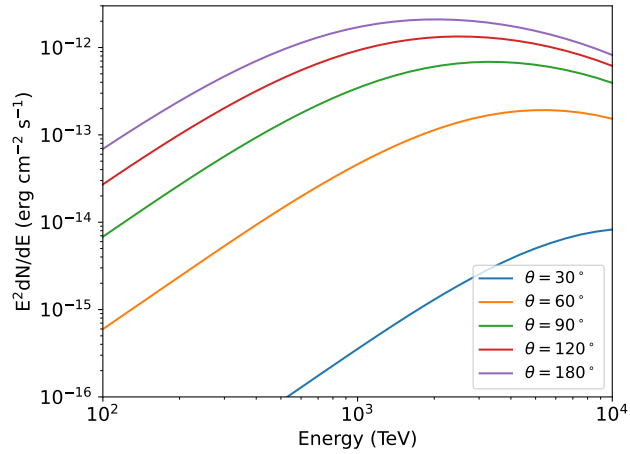

**Figure M11**: The SED of $p\gamma_{UV}$ interaction with $\alpha = 2.2$ but different interaction angles.


\end{document}